# On_the_implementation_of_the_conditions_of_Inertial_Confinement_Fusion by bombarding the target a macro particle

*S. N. Dolya*

*Joint Institute for Nuclear Research, Joliot Curie str. 6, Dubna, Russia, 141980*

The acceleration of lithium tube segments with the length $l_s = 1$ cm, diameter $d_s = 16$ μ, wall thickness $\delta_s = 1$ nm up to the energy $W_{fin} = 1$ MeV / nucl is considered. These segments are electrically charged up to the surface field strength $E_s = 10^9$ V / cm by proton beams produced by an electron beam source, which results in a charge-to-mass ratio $Z / A \approx 1.6 * 10^{-2}$. Then, they are accelerated by the traveling wave field in a spiral waveguide at the length of acceleration $L_{acc} \approx 30$ m. The segments are next sent to a frozen (D, T) target where they are compressed by three hundred times in the longitudinal direction while compressing the (D, T) target radially by $10^4$ times—thus, the conditions for thermonuclear fusion are realized.

## 1. Introduction

A prerequisite of thermonuclear burning in a reactor with inertial-confinement fusion (ICF) is the target compression by a power flux from external radiation sources. This is done in order to increase by thousands times the reaction yield $Y = \sigma n l_t$, where σ is the reaction cross section; n is the density of the target atoms; $l_t$ is the interaction length which is about the target size. While compressing a spherical target, its density increases as $1/r^3$, and the interaction length decreases as r, so that the product $n l_t$ may grow significantly, and the yield will increase from the normal for nuclear reactions values $Y \sim 10^{-4}$ to the value $Y \sim 1$ as the thermonuclear fuel is burned completely.

The well-known schemes of compression by laser radiation and light ion beams do not allow (under focusing conditions) compression of targets with a characteristic transverse dimension of several microns. However, no such limitation exists for the compression of targets by macro particles accelerated to energies above the threshold of nuclear reactions.

This study examines the acceleration of macro particles in a spiral waveguide and processes occuring as they hit a (D, T) target.

## 2. The parameters of an accelerated particle

Let us consider a macro particle representing a thin-wall lithium tube section, namely, a tube segment with the diameter $d_s = 16$ μ, wall thickness $\delta_s = 10^{-7}$ cm, (1 nm), and length $l_s = 1$ cm. The mass of the tube section is:

$M_s = \rho_{Li} * \pi \, d_s \, \delta_s \, l_s \approx 2.6 * 10^{-10}$ g, where the density of lithium is taken to be $\rho_{Li} = 0.53$ g/cm$^3$.

We find from Avogadro's ratio that the number of atoms per cubic centimeter of lithium is $4.5 * 10^{22}$ atoms. A tube segment with the specified parameters contains $N_{Li} = 2.25 * 10^{13}$ atoms of lithium or approximately $A = 1.6 * 10^{14}$ nucleons.

When irradiating such a macro particle by an ion beam, it can be electrically charged up to the surface field strength $E_s = 10^9$ V / cm. To be specific, we will refer here to a proton beam.

The density of the field electron emission current is $j$ (A/cm$^2$) = $10^{-4}$ for the surface barrier $e\varphi = 2.38$ eV (electronic work function for lithium, [1], p. 444) and the surface field intensity $E_s = 10$ MV / cm, [1], p. 461.

Let us assume that the surface intensity of the electric field on a tube segment is $E_s = 10^9$ V / cm, and the density of the leakage current of positive charges from the tube section equals $j = 10^{-4}$ A/cm$^2$.

The charge density κ located on the cylindrical surface is related to the surface intensity of the field $E_s$ by the relationship:

$$E_s = 4\kappa/d_s, \qquad (1)$$

does not depend on the cylindrical surface length, and is determined only by its diameter.

We find from (1) that at the surface field strength $E_s = 10^9$ V/cm one-centimeter of the tube section length contains $N_p = 2.6 * 10^{12}$ uncompensated protons.

Let us calculate the ratio of charge per nucleon in such a cylinder. To do so, we divide the total number of excess protons, $N_p = 2.6 * 10^{12}$, placed on a tube segment by the total number of nucleons contained in this segment, $A = 1.6 * 10^{14}$. We obtain that this ratio $N_p / A = 1.6 * 10^{-2}$, which roughly corresponds to a fourfold ionized ion of uranium 238 (4/238 = $1.7 * 10^{-2}$).

The energy of ions irradiating the macro particle should vary from the primary, equal to units of electron-volts, to the end one:

eφ = (½) $E_s$ * $d_s$ = 800 keV.

A short explanation should be given as to how such a lithium thin-wall tube segment can be made. It can be produced by spraying a few layers of lithium (10 Angstroms = 1 nm) onto the inner surface of any tube-shaped shell, for instance, from an organic material. Obviously, before such a tube segment is further irradiated to provide its positive charge, this shell should be removed, for instance, through evaporation under laser irradiation.

The head and tail parts of the tube section should be closed with lithium hemispheres of the same wall thickness with the result that it must have the shape of a vessel operating under pressure.

This segment of thin-wall lithium tube with closed hemispheres of the head and tail parts is, in fact, a lithium bubble.

**3. Acceleration of particles**

We consider the acceleration of macro particles which can be carried out according to the usual scheme: preliminary electrostatic acceleration up to a speed that approximately coincides with the phase velocity of the wave in a slowing structure and final acceleration in the traveling wave field.

*3. 1. Static field acceleration*

Let us assume that the voltage under which the platform with the container comprising tube segments is kept equals U = 800 kV. Then, after electrostatic acceleration the relative speed of the segment, β = V / c, expressed in terms of the speed of light in vacuum c = 3 x $10^{10}$ cm / s, will be:

$$\beta = [2e (Z / A) U/Mc^2]^{1/2}, \qquad (2)$$

where $Mc^2$ = 1 GeV is the rest mass of the nucleon expressed in units of electron-volt; Z / A = 1.6 * $10^{-2}$ is the electric charge per nucleon; and e is the unit charge. Substituting the numbers into formula (2), we find that the speed of macro particles after preliminary acceleration equals β = 5 * $10^{-3}$. The initial velocity $β_{in}$ of the wave traveling in a slowing structure—a wave whose longitudinal electric field component is bound to accelerate macro particles—should be the same.

*3. 2. Electromagnetic acceleration*

*3. 2. 1. The parameters of a spiral waveguide*

We choose a spiral waveguide as an accelerating structure for the acceleration of macro particles, [2].

For a spiral waveguide with the spiral winding radius $r_0 = 1$ cm, the relationship β between the winding pitch and phase velocity therein can be obtained:

$$\beta = \sqrt{2} * \mathrm{tg}\Psi/\varepsilon^{1/2}, \qquad (3)$$

where $\mathrm{tg}\Psi = h/2\pi r_0$ is the tangent of the winding angle; h is the spiral pitch; $2\pi r_0$ is the turn perimeter; ε is the dielectric permittivity factor of the medium filling the space between the spiral and the outer shield.

Let us take water as a filler of the space between the spiral and the outer shield: $\varepsilon = 80$, $\varepsilon^{1/2} \approx 9$. Then, for the spiral radius $r_0 = 1$ cm and the initial velocity of the wave in the slowing structure $\beta = \beta_{in} = 5 * 10^{-3}$, we find from relation (3) that $h_{in} = 0.28$ cm.

In a spiral, waves of different frequencies propagate with the same deceleration, but there is a frequency $f_0$ for which the maximum strength of the longitudinal field $E_z$ can be achieved at a given flow rate. This frequency may be determined from the formula:

$$2\pi r_0/\beta\lambda_0 = 1, \qquad (4)$$

where $\lambda_0 = c/f_0$ is the vacuum wavelength. From (4) we find the optimal wavelength of acceleration (for the beginning of the accelerator): $\lambda_{0in} = 12.5$ m. The corresponding frequency $f_{0in} = c/\lambda_{0in} = 24$ MHz.

We now choose an end-point energy up to which such lithium macro particles will be accelerated: $W_{fin} = 1$ MeV / nucl. Let us assume that the voltage to accelerate a micro particle $E_0 \sin\varphi_s = 20$ kV / cm, where $E_0 = 30$ kV / cm is the field amplitude; $\varphi_s = 45^0$ is the synchronous phase, $\sin\varphi_s = 0.7$.

Knowing the effective charge of a macro particle $Z/A = 1.6 * 10^{-2}$ and intensity of the accelerating field $E_0 \sin\varphi_s = 20$ kV / cm, you can find the accelerator length $L_{acc2}$ from the relationship:

$$W_{fin} = (Z/A)\, e\, E_0 \sin\varphi_s * L_{acc2}. \qquad (5)$$

Hence: $L_{acc2} = W_{fin} / [(Z/A)\, e\, E_0 \sin\varphi_s] \approx 30$ m.

Such a large accelerator length is due to the fact that the amount of charge per nucleon has a very low value for a macro particle, $(Z/A) = 1.6 * 10^{-2}$, while being approximately $(Z/A) = ½$ for light ions, and for protons equal to unity.

### *3. 2. 2. Power consumption*

We shall find the power required to create in a spiral waveguide the field strength $E_0 = 30$ kV / cm for the start of the accelerator from the relationship, [2]:

$$P_{in} = (c/8)\, E_0^2 * r_0^2 * \beta_{in} * \{(1 + I_0 K_1 / I_1 K_0)(I_1^2 - I_0 I_2) + \varepsilon (I_0 / K_0)^2 (1 + I_1 K_0 / I_0 K_1)(K_0 K_2 - K_1^2)\}, \qquad (6)$$

where $I_0, I_1, I_2$ are modified Bessel functions of the first kind, and $K_0, K_1, K_2$ are modified Bessel functions of the second kind. The first term in the curly brackets corresponds to the flux propagating inside the spiral, while the second term to the flux traveling between the spiral and external shield. The latter area is filled with a medium having a dielectric constant $\varepsilon$; therefore, the second summand contains a factor $\varepsilon$.

The Bessel functions in (6) are dependent on the argument $x = 2\pi r_0 / \beta\lambda_0$. Where the spiral is most effective, $x = 1$, the first term in the curly brackets is equal to unity and the second term to $4\varepsilon$.

In our case, $\varepsilon = 80$, the first term can be neglected, and substituting the numbers into formula (6), we obtain:

$$P_{in} = 3 * 10^{10} * 9 * 10^8 * 5 * 10^{-3} * 4 * 80 / (8 * 9 * 10^4 * 10^7) = 6 \text{ MW}.$$

Let us determine the required power for the end of the accelerator, assuming that the same spiral radius $r_0 = 1$ cm is taken. Since the phase velocity has increased by two orders towards the end of the accelerator, there is no water filling of the space between the spiral and the outer shield. Then the power $P_{fin}$ equals:

$$P_{fin} = 3 * 10^{10} * 9 * 10^8 * 4.5 * 10^{-2} * 4 / (8 * 9 * 10^4 * 10^7) = 675 \text{ kW}.$$

We can see that obtaining such power for the pulsed operation of radio frequency generators will not be a problem.

The required high-frequency power at the accelerator's beginning is greater compared to its end, which is due to the use of a dielectric between the spiral and outer shield. According to formula (6), only 0.3% of high-frequency power propagates along the spiral axis where the lithium tube segments are accelerated. This use of the dielectric is explained by the necessity to increase the spiral pitch at the beginning of the accelerator. Let us find the spiral pitch $h_{fin}$ at the end of the accelerator from the relation $\beta_{fin} = h_{fin}/2\pi r_0$. Substituting $r_0 = 1$ cm and $h_{fin} = 2.8$ mm, we obtain: $\beta_{fin} = 4.5 * 10^{-2}$.

### *3. 2. 3. The capture of macro particles into the acceleration mode. Tolerances.*

We calculate now the accuracy which is required for the initial phase of the accelerating wave to coincide with the synchronous phase. The theory of particle capture by a traveling wave shows [3]: $\Delta\varphi = 3\varphi_s$, $(+ \varphi_s, - 2\varphi_s)$. In our case, the quarter period $1/4f_0 = T_0/4$ corresponds to 10 ns or $90^0$; hence, one degree in phase corresponds to a time interval of about 0.1 ns.

A buncher in a linear accelerator provides a bunch phase width of $\pm 15^0$. In order to avoid large phase fluctuations, the accuracy of synchronization between the macro particle and accelerating wave should be:
$\Delta\tau = \pm 15 * 0.1 \text{ns} = \pm 1.5$ ns.

Such timing accuracy is apparently quite achievable.

Let us calculate the tolerance for the accuracy of matching the initial speed of a macro particle and phase velocity of the pulse propagating along the spiral structure. We introduce the value $g = (p-p_s)/p_s$, which is a relative difference in pulses, [3]. In a non-relativistic case, this corresponds to a relative velocity spread: $g = (V-V_s)/V_s$. The vertical spread of the separatrix is calculated by the formula, [3]:

$$g_{max} = \pm\, 2\, [(W_\lambda ctg\varphi_s/2\pi\beta_s) * (1 - \varphi_s / ctg\varphi_s)]^{1/2}, \qquad (7)$$

where $\varphi_s = 45^0 = \pi / 4$, $ctg\varphi_s = 1$, $[1 - \varphi_s / ctg\varphi_s]^{1/2} = 0.46$, $2 * 0.46 = 0.9$
$W_\lambda = (Z / A)\, eE_0\lambda_0 \sin\varphi_s/Mc^2$.

Further, we determine the value $W_\lambda = (Z / A)\, eE_0\lambda_0 \sin\varphi_s/Mc^2$, which is a relative energy gain by a macro particle at the wavelength $\lambda_0$ in vacuum. In our case: $\lambda_0 = c/f_0 = 12.5$ m, $\sin\varphi_s = 0.7$, $Mc^2 = 1$ GeV, $W_\lambda = 4 * 10^{-4}$. Substituting numerical values, we obtain that $g = (V_{in}-V_s) / V_s = \Delta V / V_s$, and, finally, $\Delta V/V_s = \pm\, [4*10^{-4}/ (6.28*5*10^{-3})]^{1/2}*0.9 = \pm\, 0.1$.

Thus, the admissible discrepancy between the initial speed of the macro particle and the wave velocity is of the order: $\Delta V / V_s = \pm\, 10\%$.

*3. 2. 4. The heating of the spiral*

In the above, we have found high-frequency power losses associated with attenuation. These losses are spent for heating of the spiral, and in order to prevent changing of its electromagnetic characteristics and melting, the spiral needs to be cooled. This can be done with water, which acts as a medium with a large dielectric constant and is located between the spiral and external shield.

At the end of the accelerator where there is no such medium, cooling needs to be done by gas such as helium.

*3. 2. 5. Radial movement*

As is well known, [3], in azimuth-symmetric waves the phase stability (or autophasing) region is characterized by radial defocusing. Therefore, you cannot obtain radial and phase stability in such waves concurrently. However, when the phase stability is achieved, radial stability can be provided as well by introducing external fields. In this phase region, the radial component of the electric field of the wave is directed towards radius increase, accelerating thus the particles in the direction away from the axis of acceleration.

Let us consider focusing of accelerated lithium tube segments using quadrupole lenses. It is known, [3], that quadrupole lenses simultaneously provide focusing of particles in one plane and defocusing in another. If two lenses are used so that they are turned relative to each other through $90^0$, then focusing and defocusing areas will be generated alternately in each of the transverse planes. Under certain conditions, such a system of lenses appears to be focusing one. The doublets are usually placed between adjacent sections being turned through $90^0$.

In fact, a particle moving accurately along the axis is not acted on by any forces. The further away the particle is from the axis, the greater is the action of the forces. Let the particle hit the focusing area. Its trajectory will bend then in such a way as to cross the defocusing area at the minimal value of the field, and the focusing forces will prove to be greater than defocusing ones. A similar effect also arises if the particle passes the defocusing area first. The resulting effect from a pair of quadrupole lenses will be collecting, [3].

The focusing and defocusing effects of the lenses are determined by their rigidity:

$$K = [(Z/A) eGl_l^2/Mc^2\beta_z], \qquad (8)$$

where $(Z/A)$ is the ratio of charge to mass; G is the gradient of the electric or magnetic field in the lens; $l_l$ is the length of the lens; $Mc^2 = 1$ GeV corresponds to the rest mass of the nucleon; $\beta_z$ is the longitudinal particle velocity expressed in terms of the speed of light. Rigidity among these notations is a dimensionless quantity.

In contrast to a pair of quadrupole lenses, the accelerating section operates as a defocusing lens in both perpendicular directions. Around the acceleration axis where there is no electric volume charge the following condition is realized:

$$\text{div } E = 0, \qquad (9)$$

from whence follows the ratio between the longitudinal and transverse electric fields:

$$E_r = -(r/2) dE_z/dz, \qquad (10)$$

which is apparent, however, from the structure of the field in the spiral waveguide, [2].

By analogy with quadrupole lenses, we can introduce for the accelerating field the concept of the field gradient $G_s = \frac{1}{2} dE_z / dz = \pi E_0/\lambda_s$, where $E_0$ designates the amplitude of the accelerating field; $\lambda_s$ is the length of the slow wave in the structure.

Let us consider now the focusing by electrostatic quadrupole lenses. We require that the rigidity of the quadrupole lens be greater than the rigidity of the accelerating section, which means that the particle deflection angle in the lens must be greater than the deflection angle in the section. In the accelerating section, the angle of particle deflection is always directed outwards and defocusing of the accelerated particles occurs in both transverse planes. The particles passing through the quadrupole lens are deflected inwards while being deflected outwards in a different plane. [2].

This means that the gradient of the field in the lens multiplied by the square of its length must be greater than the gradient of the field in the accelerating section, also multiplied by the square of its length:

$$G_l * l_l^2 > G_s * l_s^2. \qquad (11)$$

The most difficult conditions in terms of focusing are found at the beginning of the accelerator.

Let the length of the lens make up one-third of the section length, that is $l_s^2/l_l^2 = 10$, $l_s = 3l_l$. Then, the electric field gradient in the electrostatic lenses must exceed at least by ten times the electric field gradient in the sections, i.e:

$$G_l > 10\, G_s. \qquad (12)$$

Substituting the numbers for the start of acceleration, where $\lambda_s = 6.28$ cm, we find that the electric field gradient in the electrostatic lenses must be greater than: $G_l > 1.5 * 10^5$ V/cm$^2$.

Such high gradients in electrostatic quadrupole lenses are the "payment" for the high and continuous (for a traveling wave in lengthy sections) rate of acceleration. If such field gradients in the lenses are difficult enough to obtain, you will need to move to a lower rate of acceleration.

It needs to be added that such deceleration rates as $\beta_{in} = 5 * 10^{-3}$ can be easily achieved in accelerators with an azimuthally asymmetric accelerating field, the so-called RFQ (Radio Frequency Quadrupole) accelerators where focusing is performed by the accelerating field and no other focusing elements are necessary.

**4. The power transmitted to the beam**

We have calculated the RF power of generators needed to generate an electric field of intensity $E_0 = 30$ kV / cm in a spiral waveguide. Let us now estimate the power transferred to the beam.

Each macro particle contains $N_{Li} = 2.25 * 10^{13}$ lithium atoms with the energy $W_{Li} = 7$ MeV. Multiplying these figures, we find the energy of a macro particle in electron volts; and dividing the resulting expression by $6.24 * 10^{18}$, we obtain the energy of a macro particle in Joules: $\mathcal{E}_{Li} = 25$ J.

During the pulse operation of high-frequency generators the number of such particles in the wavetrain is $2 * 10^3$ at the pulse duration $\tau_{pulse} = 100$ µs, so that the power delivered to the beam from a generator is found to be $P_{beam} = 500$ MW, which indicates that the high frequency power of the generators must be increased, by about 100 times at the beginning of the accelerator.

It can be seen that the energy transferred to the beam is significantly greater than the energy expended on the generation of the accelerating field. If high-frequency energy is recuperated, the efficiency of the accelerator can be raised even higher.

Such power transmitted to the beam, $P_{beam} = 500$ MW, is too great for the modern high-frequency generators.

Standing-wave linear (i.e. RFQ) accelerators can be operated using stored energy. In order to reduce the power transmitted to the beam up to 5 MW a lithium tube segment needs to be placed into every hundredth separatrix.

In traveling-wave linear accelerators, the acceleration of particles can be provided by a faster running powerful single pulse instead of a high frequency wave [2], which ensures a corresponding growth of power.

Let us analyze the sine-shaped pulse with the help of the Fourier integral: $E = E_0 \sin(2\pi/T_0)t$, where $2\pi/T_0 = \omega_0$, $\omega_0 = 2\pi f_0$. We obtain:

$$f_1(\omega) = (2/\pi)^{1/2} \int_0^{T_0/2} \sin\omega_0 t * \sin\omega t \, dt. \qquad (13)$$

   The pulse spectrum is narrow and occupies a frequency range from 0 to $2\omega_0$. The dispersion (dependence of the phase velocity on the frequency) in a spiral waveguide is weak, and the waves within this frequency range are expected to propagate with the same phase velocity. As a result, the half-wave sinusoidal pulse will spread out in space due to increasing phase velocity of the wave. Matching the spiral waveguide with the supply system must be implemented in this case in the frequency range: $\Delta f \approx \omega_0/2\pi$.

   We introduce the concept of pulse amplitude $U_{acc}$ associated with the field on the spiral axis $E_{0pulse}$ using relation [2]:

$$U_{acc} = E_{0pulse}\lambda_{slow}/2\pi, \quad \lambda_{slow} = \beta\lambda_0, \quad \lambda_0 = c/f_0. \qquad (14)$$

   As a result, the amplitude of the voltage pulse propagating along the spiral axis is as follows: $U_{acc} = E_{0pulse} * \lambda_{slow}/2\pi = 30$ kV. The amplitude of the current pulse $I_{acc} = P / U_{acc} = 17$ kA; the impedance line $\rho_{line} = U_{acc} / I_{acc} = 2$ Ohms.

   Such a low resistance can be obtained in an artificial transmission-line with lumped parameters where the spiral winding appears as a distributed inductance (640 nH / cm) - the winding must be respectively
capacity - loaded (160 nF / cm).

   Thus, we have obtained a multi-section coil gun. The difference between ours and the known samples is that we are dealing with the acceleration of particles having characteristic transverse dimensions of the order of several microns. Furthermore, we have chosen the particles in the form of a bubble where the entire mass is concentrated near the surface, and to reduce the field emission we have charged the bubble with positive particles. As is known, the field ion emission arises when the field intensity begins to significantly exceed the field electron emission. This allows one to increase the charge-to-mass ratio of the particles and accelerate them to the energy above the threshold of nuclear reactions at a length of some tens of meters.

Table of parameters

| | |
|---|---|
| The accelerated lithium tube segment: length, diameter, wall thickness | $l_s = 1$ cm, $d_s = 16$ μ, $\delta_s = 1$ nm |
| The number of atoms in the segment, the number of nucleons | $N_{Li} = 2.25*10^{13}$, $A = 1.6*10^{14}$ |
| The number of protons placed on the segment | $N_p = 2.6*10^{12}$ |
| The charge per nucleon in the interval | $Z/A = 1.6*10^{-2}$ |
| The energy of protons irradiating the cylinder, initial—final | $e\varphi = 10$ eV - 800 keV |
| The potential difference relative to the ground beneath which is a high-voltage platform | $U = 800$ kV |
| The relative velocity of the segments after the electrostatic acceleration, the initial speed of the electrodynamic acceleration | $V_{in}/c = \beta_{in} = 5*10^{-3}$ |
| The radius of the spiral at the beginning and end of the acceleration | $r_{0in} = r_{0fin} = 1$ cm |
| The pitch of the spiral at the beginning and end of the acceleration | $h_{in} = h_{fin} = 2.8$ mm |
| The frequency of acceleration at the beginning and end of the acceleration | $f_{in} = 24$ MHz, $f_{fin} = 215$ MHz |
| High-frequency power spent on the generation of electric field in the structure at the beginning and end of the accelerator | $P_{in} = 6$ MW, $P_{fin} = 615$ kW |
| The relative end speed of the lithium tube segments | $V_{fin}/c = \beta_{in} = 4.5*10^{-2}$ |
| The acceleration length | $L_{acc} = 30$ m |
| The power transmitted to the beam | $P_{beam} = 500$ MW |
| The energy of each tube segment | $\mathcal{E}_{Li} = 25$ J |

Figure 1 shows a scheme of the device.

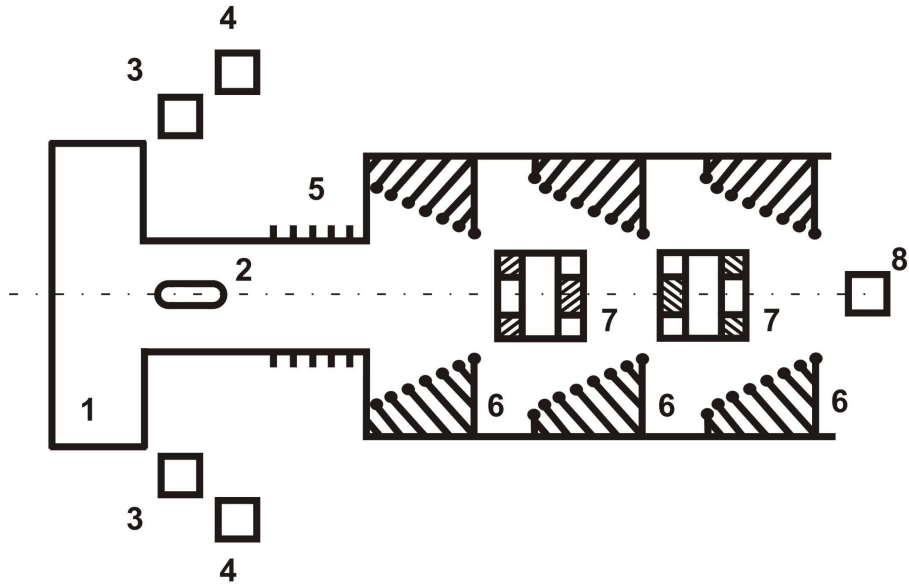

Fig.1. (1)– a container, (2) - lithium tube segments, (3) – a system of lasers, (4) - electron-beam sources, (5) – a high-voltage tube, (6) - sections of the spiral waveguide, (7) – quadrupole doublet lenses, (8) - deuterium-tritium target.

**5. Thermonuclear reactions**

Let us consider nuclear reactions which are to occur when such a lithium "bubble" will hit, for example, a deuterium-tritium target.

The density of liquid hydrogen at the temperature -260 $^0$ C is $\rho = 0.076$ g/cm$^3$; the density of deuterium is twofold higher; and of tritium three times greater, [1], p. 57. A mixture of deuterium and tritium, 50% to 50%, will have a molecular weight of 5 g and density approximately equal to 0.2 g/cm$^3$.

We find the number of molecules contained in one cubic centimeter from Avogadro's relationship whence one cubic centimeter contains 2.4 * 10$^{22}$ molecules or about 5 * 10$^{22}$ atoms.

Next, we find the depth of the path of lithium nuclei with an energy of 1 MeV / nucleon in a deuterium-tritium mixture.

The stopping power of protons in the air, at an energy of protons 1 MeV, is equal to $dW / dx = 150$ MeV * cm$^2$ / g, [1], p. 953.

In most of the path the losses are equal to $dW/dx = 600$ MeV * $cm^2/g$.

The stopping power of hydrogen in air is about two and a half times greater, [1], p. 953, and fully ionized (due to stripping) lithium ions will have the same path as protons. Altogether, one can assume that the stopping power of the deuterium-tritium mixture for fully ionized lithium ions equals: $dW/dx = 1.5$ GeV * $cm^2/g$. Multiplying the stopping power by the density of the mixture, we obtain the energy losses through ionization $dW/dx = 300$ MeV/cm and the range of the lithium ions $l_{fr}$: $W_{fin}/(dW/dx) = l_{fr} = 30$ μ.

We now find the number of atoms inside the volume bounded by the diameter of the lithium bubble, $d_s = 16$ μ, and by the length of the stopping path of lithium ions in a deuterium-tritium target, $l_{fr} = 30$ μ. The volume occupied by this tube segment is $V_b = \pi r_s^2 l_{fr} = 6 * 10^{-9}$ $cm^3$ and contains $N_{(D, T)} = 3 * 10^{14}$ atoms.

Next, we find the total energy introduced by the bubble (without the energy released by nuclear reactions, as well as taking into account that only half of the energy released goes inside): $W_{total} = (½) * 7$ MeV $* 2.25 * 10^{13} = 7.8*10^{19}$ eV. Dividing this energy by the total number of atoms in the volume bounded by the cylinder diameter $d_s = 16$ μ and by the length equal to the depth of cylinder penetration into the deuterium-tritium mixture $l_{fr} = 30$ μ, ($N_{(D, T)} = 3 * 10^{14}$), we determine that for each atom of the mixture there is approximately 260 keV energy.

The optimal energy near which the reaction yield is maximal equals 107 keV, the corresponding cross section being 5 barns, [1], p. 899.

Let the reaction yield Y be equal to the value $Y = \sigma n l_t$, where σ is the reaction cross section; n is the density of the target atoms; $l_t$ is the mean path of the incident nucleus in the target. The yield is the probability of a nuclear reaction per one incident nucleus. We take the cross section of the reaction D + T = He + n equal to 2 barns. Then, for the probability of the reaction Y to be equal to Y = 1, it is required that the product $nl_t$ be equal to $nl_t = 5 * 10^{23}$.

In our case the density of the target atoms $n_{(D, T)} = 5 * 10^{22}$, and the yield probability is unity at the interaction length $l_t = 10$ cm. It can be seen that the interaction length in this case is $l_t = 10$ μ, that is, four orders are lacking for the reaction yield to be 100%.

If the interaction of a lithium tube with the (D, T) target results in a radial compression of the target by $10^4$ times, then the density will rise by 8 orders of magnitude; the interaction length will reduce by 4 orders; and the product $nl_t$ will reach the value $nl_t = 5 * 10^{23}$.

The speed of the mixture deuterons at the energy $W_D = 260$ keV per one deuteron is: $\beta_D = V_D / c = (2W_D/Mc^2)^{1/2}$. For the deuteron: $\beta_D = 2.3 * 10^{-2}$ and $V_D = 7 * 10^8$ cm / s. The transverse dimension of the mixture after the compression by $10^4$ times is $l_t = 10^{-7}$ cm, so the lifetime of plasma in a compressed state $\tau$ can be considered equal to: $\tau = l_t / V_D = 1.4 * 10^{-16}$ s.

The initial density of the mixture $n_{in} = 5 * 10^{22}$. After a radial compression by $10^4$ times the density will increase up to $5 * 10^{30}$ atoms/cm$^3$, and the product $n\tau$ will be equal to $n\tau = 7 * 10^{14}$ atoms * s/cm$^3$, so the Lawson criterion is performed in this case. It should be kept in mind that we have assumed the energy per one atom to be equal to $W_D = 260$ keV, instead of 10 keV for which the Lawson criterion is formulated and where the reaction cross section is a thousand times smaller than at the maximum.

**6. Conclusion**

The total energy release in the (D, T) reaction is 17.6 MeV. A 100% interaction of deuterium and tritium nuclei in the target ensures the energy release $\Delta W = 17.6 * 10^6 * 1.5 * 10^{14} = 3.5*10^{21}$ eV or approximately 560 J. Assuming that one gram of TNT releases 4 kJ of energy, the energy of a microburst in this case will be equal to 140 mg of TNT.

When the accelerator is operated in the continuous mode, $f_{0in} = 2.4 * 10^7$ tube segments can be accelerated per one second. If each segment releases 560 J of energy during the interaction with the deuterium-tritium target, the total energy release will be greater than 13 GJ / s or 13 GW.

The acceleration of the cone might provide best results for the compression of the (D, T) target in comparison with the compression of the tube segment. The cone needs to be accelerated with its base forward.

References

1. Tables of Physical Data. The Handbook, edited by I.K. Kikoin, Moscow, Atomizdat, 1976

2. S.N. Dolya, K.A. Reshetnikova, About the Electrodynamic Acceleration of Macroscopic Particles, JINR Communication P9-2009-110, Dubna, 2009,
http://www1.jinr.ru/Preprints/2009/110(P9-2009-110).pdf
http://arxiv.org/ftp/arxiv/papers/0908/0908.0795.pdf

3. I.M. Kapchinsky, Particle Dynamics in Linear Resonance Accelerators, Moscow, Atomizdat, 1966